# Magnetically Levitated Microrobotic Mixer

Ecenur Can Yılmaz
*Yildiz Technical University*
Istanbul, Turkey

Abdurrahim Yılmaz
*Yildiz Technical University*
Istanbul, Turkey

Ali Anıl Demirçalı
*Yildiz Technical University*
Istanbul, Turkey

Efehan Topçu
*SEV American College*
Istanbul, Turkey

Lila Kaman
*Haileybury College*
Hertford, England UK

Hüseyin Üvet
*Yildiz Technical University*
Istanbul, Turkey

*Abstract*—Microfluidic systems, when combined with microrobots, offer enhanced precision in chemical synthesis by precisely controlling reaction conditions. These systems, when integrated with analytical tools, allow for real-time monitoring and are cost-efficient due to their minimal volume requirements, thereby reducing risks associated with hazardous chemicals. In our study, we have investigated the mixing efficiency of Thymolphthalein indicator with NaOH solution in a magnetically levitated microrobotic mixer. A PMMA microfluidic chip was used to transfer fluid containing two different solutions and achieve fast and efficient mixing. By adjusting five different flow rates and altering the rotational speeds of the microrobots, the mixing efficiency was observed. The studies were carried out under the laminar regime, with incompressible Newtonian flow rates and varying actuator speeds. The measurement of mixing efficiency was accomplished through the calculation of changes in pixel intensity observed in microscopic images acquired throughout the mixing process. The presence of the microrobots resulted in the best efficiency at 80.37% at 500 rpm and 7 $mL/min$ flow rate. Their potential in advanced reactions, such as nanoparticle synthesis and encapsulation, suggests promising avenues for improving product yields.

## I. Introduction

Microfluidics, an interdisciplinary field at the intersection of engineering, chemistry, and biology, has emerged as a transformative technology with far-reaching implications across a multitude of scientific and industrial applications. From medical diagnostics and drug delivery systems to chemical microreactors and process engineering, microfluidic devices have revolutionized our approach to fluid control and manipulation [1]. These devices are characterized by their ability to handle minuscule volumes of fluids, offering rapid, cost-effective, and precise analyses with minimal sample requirements. Moreover, they provide unparalleled advantages in terms of scalability, precision, and real-time monitoring capabilities, setting them apart from traditional macro-scale systems [2].

Microfluidic platforms have emerged as versatile tools, finding applications in a range of configurations such as microvalves, micropumps, microchannels, and micromixers. These configurations are made possible due to the advancements in production techniques tailored for microscale operations. Among the various components, micromixers are serving as the heart of the system to ensure thorough and efficient mixing of solutions. The significance of such microfluidic mixing processes cannot be overstated, especially when considering their transformative impact on the biological and chemical industries, where precision and efficiency are paramount. Madadelahi et al. presented a comprehensive review on the mathematical modeling and computational analysis of centrifugal microfluidic platforms, emphasizing their potential in automating chemical and biological assays [3]. Timilsina et al. embarked on a journey to understand the biological and chemical properties of the essential oil and extracts of the rhizome of Acorus calamus Linn, showcasing the potential of microfluidic techniques in extracting and analyzing natural compounds [4]. Kaminski and Garstecki delved into droplet microfluidics, emphasizing its role in studying hydrodynamics and properties of biphasic flows at the microscale. Their work highlighted the unique characteristics of microdroplets, such as rapid mixing and excellent control over reaction conditions, making them ideal for single cell or single molecule assays [5]. Saha et al. synthesized and characterized metal complexes of an ionic liquid-supported Schiff base, emphasizing the role of microfluidic systems in the synthesis and analysis of complex chemical structures [6].

However, despite these advancements, one of the most pressing and persistent challenges in the realm of microfluidics is the efficient mixing of fluids, particularly in confined geometries [7]. This challenge is exacerbated by the unique dynamics of fluid flow at the microscale, which are fundamentally different from those at the macroscale. In macroscopic systems, turbulent flow, typically associated with high Reynolds numbers, ensures rapid mixing through chaotic fluid motion. Conversely, microfluidic systems predominantly operate under low Reynolds numbers, resulting in laminar flow conditions where fluid layers slide past each other with minimal lateral mixing. This leads to a reliance on inherently slow, diffusion-based processes, which are often inadequate for applications requiring rapid and precise analyses, such as medical diagnostics and chemical synthesis [8].

To address these challenges, a significant body of research has been dedicated to the development of specialized devices known as micromixers. These devices can be broadly categorized into passive and active types. Passive micromixers, such as T- and Y-micromixers or those incorporating herringbone-inspired microstructures, exploit channel geometries and flow manipulations to enhance mixing. Innovations like the Dean micromixer have been particularly effective in disrupting laminar flow patterns, thereby promoting more efficient mixing. However, their efficacy can be limited, especially at lower flow rates where diffusion dominates [9, 10]. Active micromixers, on the other hand, utilize external energy sources like ultrasound, thermal gradients, and acoustic standing waves to augment the mixing process. These methods offer a higher degree of control but can introduce complexities, potentially compromising device integrity and increasing fabrication challenges [11].

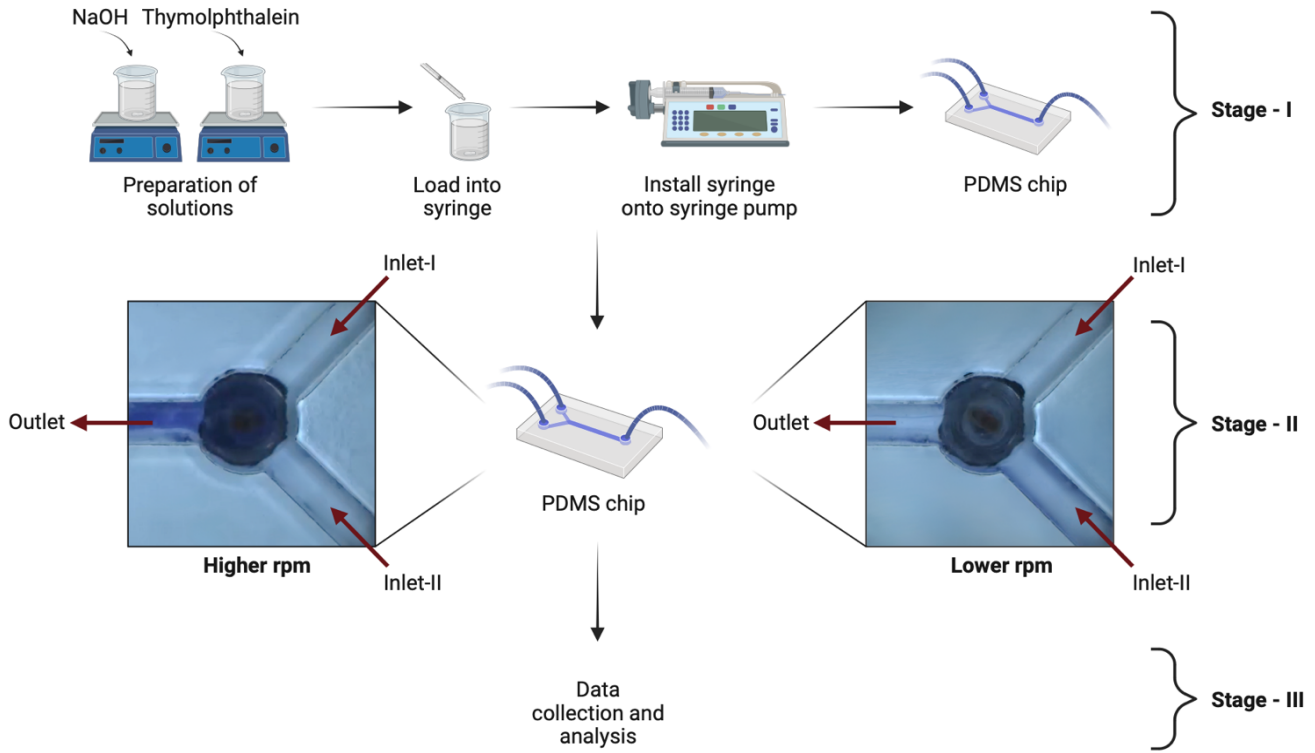

Fig. 1. The figure depicts an overview of the experimental setup. Stage-1 represents the preparation phase, while Stage-2 details the process within the microfluidic chip. The accompanying images showcase the mixture at a consistent flow rate of 5 $mL/min$, but at varying rpm values (right: 250 rpm and left: 500 rpm). Images taken during the experiment were post-processed to calculate mixing efficiency in Stage-3.

Mahmud et al. delved into the challenges of achieving rapid and efficient fluid mixing in microchannels, particularly due to the inherent laminar flow. They explored the potential of extraneously applied ultrasound and thermal energy to enhance the mixing process in passive micromixers, highlighting the benefits of such external energy sources in reducing fabrication complexities [12]. Husain et al. provided a critical overview of passive micromixers developed over the past decade, categorizing them based on design features and evaluating their mixing performance [13]. Giraldo et al. introduced a novel microfluidic device equipped with a serpentine microchannel, designed to enhance the encapsulation of nanobioconjugates within liposomes, demonstrating the versatility of micromixers in bio-applications [14].

Magnetic micromixers have garnered significant attention. These devices leverage magnetic fields to drive magnetic beads or fluids, inducing convective mixing. Their precise control and ease of integration into microfluidic systems make them a popular choice in many applications [15]. Emerging at the forefront of this quest for enhanced mixing efficiency is the integration of microrobots into microfluidic systems. With their unparalleled precision, control, and adaptability, microrobots offer a revolutionary avenue for tailoring mixing processes to specific requirements. Preliminary studies have indicated that these robotic systems can significantly enhance mixing efficiencies, especially at lower Reynolds numbers where their mechanical effects are more pronounced [16].

Building on this foundation, our research aims to delve deeper into the realm of active micromixing by focusing on a microfluidic system that leverages the capabilities of microrobots. Specifically, we will explore the mixing efficiency of Thymolphthalein, a base indicator, with NaOH solution in our experimental setup as shown in Fig. I. Through this investigation, we aspire not only to shed light on the potential of microrobots in enhancing mixing efficiency but also to provide valuable insights into the future of microfluidic systems and their transformative potential in advancing chemical synthesis and analysis at the microscale.

## II. THEORETICAL BACKGROUND

The Reynolds number acts as a key parameter for understanding the dynamics of fluid flow systems. It is a dimensionless quantity that provides insight into the relationship between the forces of inertia and viscosity within the fluid. Specifically, it offers a way to gauge the relative significance of these two forces in shaping the flow behavior. In the context of microfluidic systems, which are characterized by laminar flow, the Reynolds numbers are typically low. This low value indicates that viscous forces dominate over inertial forces in such systems. The Reynolds number is shown in the Eq. 1 where $U$ is the average flow velocity and $D_h$ is the hydraulic diameter of the channel.

$$Re = \frac{\rho U D_h}{\mu} \quad (1)$$

In our study, we focus on laminar flow conditions and consider the fluid to be incompressible and Newtonian in nature. The velocity distribution within the flow is described by the Navier-Stokes equation (Eq. 2) and the continuity equation (Eq. 3). These equations are applicable for incompressible fluid movement within the channel where $\rho$ represent fluid density, $u$ stands for the velocity vector, $t$ is the time, $p$ indicates pressure, and $\eta$ is the dynamic viscosity of the fluid.

$$\rho \cdot \left(\frac{\partial y}{\partial x} + (u \cdot \nabla)u\right) = \nabla p + \eta \nabla^2 u \qquad (2)$$

$$\nabla \cdot u = 0 \qquad (3)$$

In this research, two microrobots were employed. Their interactions are governed by magnetic forces. The mathematical representation of these magnetic interactions can be described using Maxwell's equations, which outline the nature of the magnetic fields that the microrobots generate and experience in relation to each other. The equations are as follows:

$$\nabla \cdot \vec{B} = 0 \qquad (4)$$

$$\vec{B} = \mu_0(\vec{M} + \vec{H}) \qquad (5)$$

$$\nabla \times \vec{H} = 0 \qquad (6)$$

where $\vec{H}$ is the magnetic field strength, $\vec{M}$ is the magnetization of the NdFeB magnets, $\vec{B}$ is the magnetic flux density and $\mu_0$ is the magnetic conductivity number. The force and torque applied by the actuator microrobot to the mixer microrobot are shown in Eq. 7 and Eq. 8.

$$\vec{F} = \int_V (\vec{M} \cdot \vec{\nabla})\vec{B}(x,y,z)dV \qquad (7)$$

$$\vec{T} = \int_V \vec{M} \times \vec{B}(x,y,z)dV \qquad (8)$$

where $\vec{F}$ and $\vec{T}$ are the general magnetic torques and forces, respectively. $\vec{B}(x,y,z)$ is the magnetic field at $x, y, z$ axis and $V$ is the volume of magnets.

In the existing scientific literature, a variety of mathematical equations have been employed to evaluate mixing efficiency, including those that focus on point and mean concentration. However, in the current study, a distinct methodology was adopted to quantify mixing efficiency. Specifically, the images captured during the experiment were first converted into grayscale format. This step was crucial for the subsequent analysis. It's important to note that the conditions under which these images were captured can vary, making it essential to normalize the images for a consistent analysis. While there are multiple normalization techniques that have been proposed in past research, this study established its own set of criteria for normalization. The intensity values of the fluid entering the mixing area were carefully measured, and the lowest and highest of these values were used to set the normalization limits. Once these limits were established, the mixing region within the images was normalized accordingly. Following this, the mixing efficiency was calculated by measuring the mean intensity along a straight line that extends through the channel's exit point. This calculation was considered valid only when the mixture had reached a steady state, at which point the mixing index was set to 1. The specific formula used to determine this mixing index is outlined in Equation 9.

$$MI = \sum_{i=1}^{n} \frac{C_i}{n} \qquad (9)$$

where $MI$ is the mixing index also known as mixing efficiency, $C_i$ is the concentration at the specific points, and $n$ is the number of points.

III. METHODOLOGY

*A. Materials*

A microfluidic chip was designed to function as a micro mixer in this study. This chip was fabricated from polymethyl methacrylate (PMMA) and had dimensions of 24 mm x 40 mm x 2 mm. The chip's architecture included two input channels, a single mixing chamber, and one output channel. For the mixing step, two microrobots were used. The manufacturing process for these microrobots commenced with the application of a base layer of positive photoresist (AZ1505 from Micro Chemical GmbH) onto the substrate. Subsequently, a spin coater rotating at a speed of 1100 rpm was used to apply a layer of negative adhesive film, with a thickness of 250 onto this sacrificial layer. This film was then exposed to ultraviolet light to develop a pattern, which served as the blueprint for the final polymer structure of the microrobot. Following that, three neodymium (NdFeB) permanent magnets (grade N52) were used. These magnets were carefully integrated into the polymer body of the microrobot, completing its construction and preparing it for its role in enhancing the mixing efficiency within the microfluidic chip.

NaOH solution and Thymolphthalein indicator were used as chemical reagents for the mixing process. NaOH was purchased from Sigma-Aldrich (Missouri, USA) in pellet form. Thymolphthalein indicator was purchased from Merck (Darmstadt, Germany).

*B. Experiment*

The aim of the study was to effectively blend NaOH and indicator solutions that were introduced into a microfluidic chip. The innovative robotic mixer designed for this purpose operates based on the collaboration of two microrobots. Within this system, a stepper motor functions as the main actuator. The initial microrobot is affixed to the stepper motor's rotor using double-sided adhesive (DSA) and acts as the actuating robot, driving the mixing microrobot within the microfluidic pathway.

Setting up the micro mixing system involves a specific procedure. The mixing robot is carefully placed inside the chamber of the microfluidic channel. To ensure precise alignment and proximity between the microfluidic channel and the stepper motor, a manually operated lab jack platform is employed. This step is crucial as it ensures that the magnetic fields of the two microrobots are in perfect alignment, facilitating their interaction. As a result, when the actuator motor is activated and set into motion, it allows the mixing robot to be manipulated and controlled without any physical contact.

To begin the mixing procedure, NaOH and thymolphthalein solutions were prepared. 0.1 gram of thymolphthalein is added to 50 mL of 95% ethanol solution. After dissolving, distilled water is added to complete the total volume to 100 $mL$. The approximate concentration of the prepared solution is 1 $g/L$. NaOH solution was prepared to pH 11.5, thus ensuring proper functioning of the indicator. To do this, firstly, the main stock solution is prepared by dissolving 0.277 $g$ of NaOH solid in 100 $mL$ of distilled water. The concentration of the main stock is $6.92 \times 10^{-2}$ $M$. 210 $mL$ of distilled water is added to 10 $mL$ of main stock solution. The concentration of the new solution obtained is $3.16 \times 10^{-3}$ $M$ and its pH is 11.5.

After the preparations are completed, the solutions were transferred into syringes. Two of these syringes were integrated into a single syringe pump. The system's control mechanism involved the use of an Arduino Uno type microcontroller, which was programmed to send specific signals to the stepper motor driver. This allowed for regulation of the motor's rotational speed. The mixing experiment was carried out at speeds of 0, 100, 250, 500, and 800 rpm. The syringe pump was operating at flow rates of 0.5, 1, 3, 5, and 7 $mL/min$. In total, 25 tests were performed, encompassing five distinct rotational speeds and five varied flow rates. To provide a visual representation and documentation of the entire experimental process, a high-resolution Basler acA2440-75uc camera, boasting a 5 MP resolution, was used to capture the mixing process in real-time.

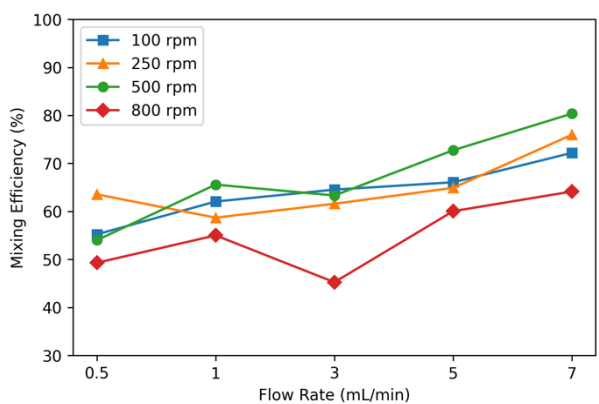

Fig. 2. The efficiency curves for 100, 250, 500, and 800 rpm at the corresponding flow rate values.

## IV. EXPERIMENTAL RESULTS

In this research, a solution of NaOH and the thymolphthalein indicator were mixed. The study aimed to evaluate the efficiency of the mixer by analyzing the outcomes at different microrobot rotation speeds and various flow rates. Once the mixture achieved a consistent and stable state, it was considered to have reached optimal mixing. This state was then documented and utilized as a benchmark or reference image. Using this image as a standard, the success and efficiency of subsequent mixing experiments were methodically assessed. The Mixing Index (MI) values, which provide a quantitative measure of the mixing efficiency, are systematically presented in Table I. In addition, efficiency curves for different rpm values are shown in Fig. II.

TABLE I
EFFICIENCY RESULTS TAKEN FROM EXPERIMENTAL SETUP

| Flow Rate \ rpm | 100 | 250 | 500 | 800 |
|---|---|---|---|---|
| 0.5 $mL/min$ | 55.23% | 63.55% | 54.01% | 49.31% |
| 1 $mL/min$ | 62.06% | 58.70% | 65.59% | 55.01% |
| 3 $mL/min$ | 64.54% | 61.60% | 63.32% | 45.26% |
| 5 $mL/min$ | 66.07% | 64.91% | 72.74% | 60.05% |
| 7 $mL/min$ | 72.20% | 75.97% | 80.37% | 64.15% |

The experimental design was both detailed and multifaceted. A series of tests spanned across five unique flow rates, each corresponding to a specific rotation speed of the microrobot. The initial phase of experimentation was understanding the inherent mixing capabilities of the solutions when subjected solely to flow, without any intervention from the microrobot. At slower flow rates, the solutions demonstrated a propensity to blend to some degree, a phenomenon attributed to the adequate rate of diffusion at these reduced speeds. An inverse relationship became evident between the speed of the flow channel and the mixing index; as the flow rate diminished, the mixing index correspondingly rose, signaling enhanced mixing. Conversely, at elevated flow rates, the inherent diffusion rate proved inadequate for complete mixing. In such instances, the solutions remained distinct, hinting at potential mixing only when they neared specific boundaries within the mixing chamber. As the flow rate escalated, the potential mixing area at these boundaries shrank. Consequently, the mixing efficiency at the exit point also saw a decline with increasing flow rates. There was no reverse diffusion observed towards the entrance channels.

In the first experimental setup, where microrobots were tasked with mixing through rotation, they operated at a speed of 100 rpm. At this rotation speed, the experiments revealed a challenge: the torque generated wasn't sufficient to effectively maneuver the mixer robot inside the chip. This inadequacy prevented the robot from achieving full levitation. As a result, the mixer robot couldn't maintain a steady rotation. At a flow rate of 0.5 $mL/min$, a Mixing Index (MI) of 55.23% was achieved. This indicated that there was a low degree of mixing in the chamber at this flow rate. However, at increased flow rates like 1, 3, 5, and 7 $mL/min$, achieving a homogeneous mix became easier as the MI exceeded 70%. Building on these findings, the research then transitioned to a second experimental phase. Here, the rotation speed of the microrobots was increased to 250 rpm. MI values of 61.60%, 64.91%, and 75.97% were recorded for flow rates of 3 $mL/min$, 5 $mL/min$, and 7 $mL/min$, respectively. These results indicate effective mixing at faster speeds with a high MI. However, the system did face challenges at slower flow rates of 0.5 $mL/min$ and 1 $mL/min$, where indications of non-uniform mixing began to surface. No homogenous mixing was observed at these flow rates. The third experimental phase the microrobots' rotation speed was increased to 500 rpm. At this speed, the MI values were similar to those observed during the 250 rpm phase. For flow rates of 0.5 $mL/min$, 1 $mL/min$, and 3 $mL/min$, the system achieved MI values of 54.01%, 65.59%, and 63.32%, respectively. At 7 $mL/min$ flow rate, the highest mixing efficiency of the system with 80.37% was observed. MI results at this 450 rpm speed were about 10% higher than those at 250 rpm, suggesting a more consistent mixing process. The fourth and final experimental phase was conducted at a microrobot at 800 rpm. It was observed that as the microrobot speed increased, the mixing efficiency decreased as the stability of the robot decreased. The system achieved MI values of 49.31%, 55.01%, 45.26%, 60.05%, and 64.15%, at flow rates of 0.5 $mL/min$, 1 $mL/min$, 3 $mL/min$, 5 $mL/min$, and 7 $mL/min$, respectively.

When analyzing the experiments across the different flow rates, it was evident that the 800 rpm speed yielded suboptimal MI values for all flow rates. In contrast, the results from 100 rpm and 250 rpm were closely aligned, with the 100 rpm speed showing a marginally higher MI value. Importantly, as the rotation speed was ramped up, especially at the higher flow rates of 5 $mL/min$, 7 $mL/min$, and 10 $mL/min$, there was a marked increase in the MI value, pointing to enhanced mixing capabilities.

## V. DISCUSSION

Over the years, the scientific community has extensively explored magnetic field-driven micro mixers, as evidenced by the plethora of research articles available in the literature. Magnetic mixers stand out due to their distinct advantages over other types of micro mixers. Notably, they remain unaffected by environmental factors such as pH and are also cost-effective. Our research introduces a magnetically levitated robotic mixer that hinges on the functionality of two microrobots. This design allows for the mixing of microfluids in a non-contact manner. Our microrobot boasts the capability of achieving high rotation speeds, and as the rotation speed of the mixer robot is ramped up, there's a corresponding increase in mixing efficiency. A notable advantage of our approach is its applicability in closed microfluidic chips. Our experiments have demonstrated that at a rotation speed of 500 rpm, we can achieve a commendable mixing efficiency of 80.37% for indicator solutions.

## VI. CONCLUSION

This work contributes to the literature by introducing a novel magnetic field drive micro mixer system. Central to this innovation is the use of a levitated micro robot, which, when spun, acts as the primary mixing agent. The system employs a mixer microrobot to propel another microrobot affixed to a stepper motor. To assess the efficacy of this novel system, a series of experiments were carried out on a Y-shaped PMMA microfluidic chip, a platform chosen for its compatibility with the proposed system. During these tests, the robot was set to rotate at speeds of 100, 250, 500, and 800 rpm, while flow rates of 0.5, 1, 3, 5, and 7 $mL/min$ were maintained. To ensure objective and quantifiable results, the entire process was microscopically documented, capturing every nuance of the mixing process. One of the standout revelations from these experiments was the system's ability to achieve a remarkable mixing efficiency. A peak efficiency of 80.37% was recorded, which is noteworthy not just for the high value but also for the fact that this was accomplished using a non-contact mechanism. The empirical data gleaned from these experiments serves as a testament to the system's viability as an effective mixing mechanism. It not only meets but, in many aspects, surpasses the benchmarks set by traditional methods. Looking ahead, future explorations will delve into the system's capability to mix nanoparticles, considering varying rotation speeds and robot designs.

REFERENCES

[1] V. Rudyak and A. Minakov, "Modeling and optimization of Y-type micromixers," *Micromachines,* vol. 5(4), pp. 886-912, 2014.

[2] C. Cierpka and C. J. Kähler, "Particle imaging techniques for volumetric three-component (3D3C) velocity measurements in microfluidics," *Journal of visualization*, vol. 15, pp. 1-31, 2012.

[3] M. Madadelahi, L. F. Acosta-Soto, S. Hosseini, S. O. Martinez-Chapa, and M. J. Madou, "Mathematical modeling and computational analysis of centrifugal microfluidic platforms: a review," *Lab on a Chip*, vol. 20, no. 8, pp. 1318-1357, 2020.

[4] R. Timilsina, P. Tandukar, and I. Pathak, "Biological and chemical studies of essential oil and extracts of rhizome of Acorus calamus Linn," *Journal of Nepal Chemical Society*, vol. 43, no. 1, pp. 35-42, 2022.

[5] T. S. Kaminski and P. Garstecki, "Controlled droplet microfluidic systems for multistep chemical and biological assays," *Chemical Society Reviews*, vol. 46, no. 20, pp. 6210-6226, 2017.

[6] S. Saha, G. Basak, and B. Sinha, "Physico-chemical characterization and biological studies of newly synthesized metal complexes of an Ionic liquid-supported Schiff base: 1-{2-[(2-hydroxy-5-bromobenzylidene) amino] ethyl}-3-ethylimidazolium tetrafluoroborate," *Journal of Chemical Sciences*, vol. 130, pp. 1-9, 2018.

[7] T. J. Ober, D. Foresti, and J. A. Lewis, "Active mixing of complex fluids at the microscale," *Proceedings of the National Academy of Sciences*, vol. 112, no. 40, pp. 12293-12298, 2015.

[8] Z. Wu, and N. T. Nguyen, "Convective–diffusive transport in parallel lamination micromixers," *Microfluidics and Nanofluidics*, vol. 1, p. 208-217, 2005.

[9] M. Bayareh, M. N. Ashani, and A. Usefian, "Active and passive micromixers: A comprehensive review," *Chemical Engineering and Processing-Process Intensification*, vol. 147, p. 107771, 2020.

[10] F. Mahmud, K. F. Tamrin, S. Mohammaddan, and N. Watanabe, "Effect of thermal energy and ultrasonication on mixing efficiency in passive micromixers," *Processes*, vol. 9(5), p. 891, 2021.

[11] C. Olivier, G. Penelet, G. Poignand, and P. Lotton, "Active control of thermoacoustic amplification in a thermo-acousto-electric engine," *Journal of Applied Physics*, vol. 115(17), 2014.

[12] F. Mahmud, K. F. Tamrin, S. Mohamaddan, and N. Watanabe, "Effect of thermal energy and ultrasonication on mixing efficiency in passive micromixers," *Processes*, vol. 9, no. 5, pp. 891, 2021.

[13] A. Husain, A. I. Khan, W. Raza, N. Al-Rawahi, N. Al-Azri, and A. Samad, "DESIGN AND MIXING PERFORMANCE OF PASSIVE MICROMIXERS: A CRITICAL REVIEW," *The Journal of Engineering Research [TJER]*, vol. 19, no. 2, pp. 106-128, 2022.

[14] K. A. Giraldo, J. S. Bermudez, C. E. Torres, L. H. Reyes, J. F. Osma, and J. C. Cruz, "Microfluidics for multiphase mixing and liposomal encapsulation of nanobioconjugates: passive vs. acoustic systems," *Fluids*, vol. 6, no. 9, pp. 309, 2021.

[15] Q. Hu, J. Guo, Z. Cao, and H. Jiang, "Asymmetrical induced charge electroosmotic flow on a herringbone floating electrode and its application in a micromixer," *Micromachines*, vol. 9(8), p. 391, 2018.

[16] J. Marschewski, S. Jung, P. Ruch, N. Prasad, S. Mazzotti, B. Michel, and D. Poulikakos, "Mixing with herringbone-inspired microstructures: Overcoming the diffusion limit in co-laminar microfluidic devices," *Lab on a Chip*, vol. 15(8), p. 1923-1933, 2015.